\documentclass[twocolumn,letterpaper,showpacs,preprintnumbers,amsmath,prb]{revtex4}
\usepackage{graphicx}
\usepackage{color}
\usepackage{verbatim}
\usepackage{hyperref}

\begin{document}
\title{Half quantum vortex in superfluid $^3$He-A phase in
parallel plate geometry}
\author{ Hae-Young Kee$^1$ and  Kazumi Maki$^2$}
\affiliation{$^1$ Department of Physics, University of Toronto,
Toronto, Ontario  M5S 1A7,  Canada \\
$^2$ Department of Physics and Astronomy, University of Southern California,
Los Angeles, CA 90089-0484,  USA }

\begin{abstract}
The half quantum vortex(HQV) in condensate has
been studied, since it was predicted by Salomaa and Volovik in superfluid $^3$He-A phase.
However, an experimental evidence for its existence has not been reported so far.
Motivated by a recent experimental report by Yamashita et al\cite{yamashita}, we study
the HQVs in superfluid $^3$He confined between two parallel plates with 
a gap D $\sim$ 10 $\mu$m in the presence of a magnetic field 
H $\sim$ 26 mT perpendicular to the parallel plates. 
We find that the bound HQVs are more stable than the singular vortices and
free pairs of HQVs, when the rotation perpendicular to the parallel plates is
below the critical speed, $\Omega_c \sim$ 2  rad/s.
The bound pair of HQVs accompanies the tilting of ${\hat d}$-vector out of the plane,
which leads to an additional absorption in NMR spectra. Our study appears
to describe the temperature and rotation dependence of the observed satellite NMR signal,
which supports the existence of the HQVs in $^3$He.
\end{abstract}
\maketitle

\section{Introduction\label{sec:intro}}
The superfluid  $^3$He-A is a remarkable condensate with several
broken symmetries and a variety of topological defects.
It was proposed that many concepts and notions generated in the high
energy particle physics can be tested in the earth bound ultra-low 
temperature laboratories.\cite{salomaa87,vollhardt90,volovik91}.
Indeed a variety of vortices, which include the continuous vortices
(the vortices with soft-core) have been observed in the rotating superfluid
$^3$He-A.\cite{hakonen86,volovik00}
All of the continuous vortices are associated with ${\hat l}$-texture
are stable in the open systems where angular momentum of Cooper pair,
${\hat l}$-vector is not constrained.

On the other hand, in the parallel plate geometry, where superfluid $^3$He
is confined between two parallel plates with a gap $D \sim \xi_D$,
where $\xi_D$ ($\sim$ 10   $\mu$m) is the dipole coherence length,
the ${\hat l}$-vector is forced to be perpendicular to 
the plate.\cite{ambegaokar74,bruinsma79}
In such a geometry, the continuous vortices cannot exist, and
one may conclude that the vortices with the circulation 
$N=\pm 1$ called the phase vortices or the singular vortices are only
possible vortices.\cite{salomaa87}
However, Salomaa and Volovik\cite{salomaa85} predicted that the half quantum
vortices(HQVs) with $N = \pm \frac{1}{2}$ should exist in the 
parallel plate geometry.
In spite of intensive experimental search, the HQV has not been found in the 
rotating superfluid $^3$He-A in the confined geometry.\cite{volovik00}

It is worthwhile to note that the HQVs have played an important role 
outside the superfluid $^3$He-A. For example, the HQVs in the tricrystal
geometry in the high temperature cuprate superconductors played the crucial
role in identifying $d$-wave symmetry of the 
high temperature superconducting order parameter.\cite{tsuei00}
The HQVs in triplet superconductors have been also suggested
\cite{sigrist99,kee00,maki06} to 
interpret the extremely strong vortex pinning observed in triplet
superconductors such as UPt$_3$, U$_{1-x}$Th$_x$Be$_{13}$, and
Sr$_2$RuO$_4$.\cite{dumont00}

The present work is  motivated by the recent experiment
by Yamashita et al.\cite{yamashita} 
We consider the superfluid $^3$He-A confined between
the two parallel plate with the gap $D \sim \xi_D$, where
the rotation axis and the magnetic field 
are perpendicular to the parallel plates. Using the standard
formula for free energies given in [\onlinecite{vollhardt90,kee00}], we found
the followings.
(a) The singular vortices with $N = \pm 1$ are unstable and split into
pairs of HQV with $N=\pm \frac{1}{2}$, unless the rotation speed
is very high, $\Omega > \Omega_u \sim 5 \times 10^6$  rad/s.
(b) For small rotation speed ($\Omega \le \Omega_c \sim$ 2 rad/s),
the bound pairs of HQVs are more stable than the free HQVs.
(c)  Only the bound pairs of HQVs are detectable by NMR as
satellite signals. Both the singular vortices and free HQVs lead to NMR signals
indistinguishable from backgrounds,
since they do not disturb the configuration of ${\hat l} \perp {\hat d}$.
(d) Our results of the satellite frequencies and the intensities are
consistent with the observation of satellite peak in the parallel
plate geometry.\cite{yamashita}
In particular, while $\Omega_c$ weakly depends on the temperature and the 
magnetic field strength, both the satellite frequency 
and the satellite intensity are almost independent of the field
strength and temperature.

\section{Free energies of one singular vortex and a pair of HQV}
The superconducting order parameter of superfluid $^3$He-A is characterized
by 2 unit vectors called l-vector, ${\hat l}$ and d-vector, ${\hat d}$.
We can write the gap function as\cite{vollhardt90,volovik91}
\begin{equation}
\Delta_{ab} ({\bf k}) = {\vec D} ({\bf k}) \cdot 
({\vec \sigma} i \sigma_2)_{ab}
\end{equation}
where 
\begin{equation}
{\vec D}({\bf k})=\Delta {\hat d} ({\hat k}_1 + i {\hat k}_2)
\end{equation}
Here $\sigma_{\mu}$ ($\mu = 1,2,3$) are Pauli matrices and $a$, $b$
are spin $\uparrow$ and $\downarrow$.
${\hat k}_i$ ($i=1,2$) are the projection of the unit wave vector ${\hat k}$
in the plane perpendicular to ${\hat l}$.
Therefore ${\hat l}$ corresponds to the quantization axis of the angular
momentum of the Cooper pair. This is also called the chiral vector.

When the spatial variation of ${\hat d}$ and ${\hat l}$ are much slower
than $\xi$ ($\sim 0.01 \mu m$), coherence length, the spatial
configuration of ${\hat l}$ and ${\hat d}$ are described
by the texture free energy.\cite{vollhardt90}.
In the following, we shall compare the texture free energy of a singular
vortex(or phase vortex), and a free pair of HQV with an  optimized distance $R$,
where $R$ is a distance between two HQVs.

The texture free energy is given by 
\begin{equation}
F= \frac{1}{2} \chi_N C^2 \int dx dy 
\left(  K (\nabla \Phi)^2 +\sum_{ij} |\partial_i {\hat d}_j|^2
+\xi_H^{-2} {\hat d}_z^2 \right),
\end{equation}
where 
we assumed the parallel plates extend in the $x-y$ plane.
Here $\chi_N$ and $C$ are spin susceptibility and spin wave velocity.
Then $K$ is given by the ratio between the superfluid density, $\rho_s$
and spin superfluid density, $\rho_{sp}$.
\begin{equation}
K= \frac{\rho_s}{\rho_{sp}} =\frac{1+\frac{1}{3} F_1}{1+\frac{1}{3} F_1^a}
\frac{1+\frac{1}{3} F_1^a (1-\rho_s^0)}{1+\frac{1}{3} F_1 (1-\rho_s^0)}
\end{equation}
where $F_1$ and $F_1^a$ are Landau's Fermi liquid coefficients.
We also assumed ${\hat l} \parallel {\hat z}$ everywhere due to
the parallel plate constraint.\cite{ambegaokar74}
Within the weak-coupling theory of BCS p-wave superconductor,
we obtain
\begin{eqnarray}
\rho_s^0 =  \left\{ \begin{array}{c}  2(1-t) \;\;\;\; {\rm for}\;\;\;\;  t \sim 1 \\
1- O(t^3)\;\;\;\;  {\rm for}\;\;\;\;  t \rightarrow 0 \end{array} \right.
\end{eqnarray}  
where $t= T/T_c$. 
For the pressure of $27.3$ bar with the superfluid transition temperature
$T_c = 2.46$ mK corresponding to the experimental set-up
by Yamashita et al\cite{yamashita}, we found
\begin{equation}
K = 1 + 2.295 (1-t),
\end{equation}
where $(1-t) << 1$.

When the parallel plates are rotated with the rotation axis perpendicular
to the plates and rotation speed $\Omega$, the density of $N=1$ vortices
is given by
\begin{equation}
n_v = (\pi a^2)^{-1} = 2 \Omega/( p \kappa)
\end{equation}
where $\kappa = 2 \pi \hbar/ (2 m_{^3He}) = 6.59 \times 10^{-8} {\rm m^2/s}$ is the quantum
of circulation, and  $p$ is the number of quanta per vortex.
 $2a$ corresponds roughly the distance between two vortices, if 
we divide the two-dimensional space in circular cells with radius $a$.
Therefore $a$ is related to $\Omega$ by $a=\sqrt{p \kappa/(2 \pi \Omega)}$.
The free energy for one singular vortex in unit cell is given by
\begin{equation}
F_s = \pi \chi_N C^2 K {\rm ln}\left(\frac{a}{\xi} \right),
\label{eq_f_singular}
\end{equation}
which does not involve the texture energy except $\Phi =
\tan^{-1}(y/x)$. Here $\xi \sim$ 0.01 $\mu$m is the 
coherence length.

\begin{figure}[t]
\includegraphics[width=2.7in,clip]{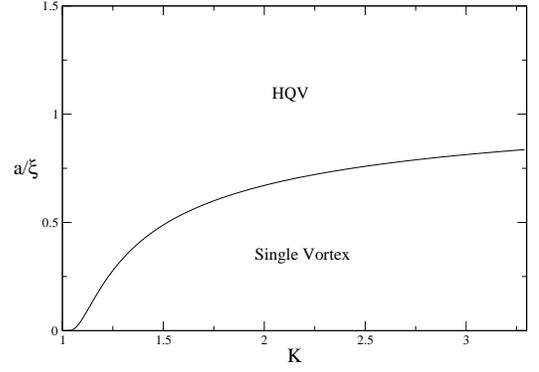}
\caption{Phase diagram for SV and free HQV phases as functions of $K$ and $a/\xi$.
Singular vortex (SV) phase is stable only for small $a/\xi$ for all temperatures.}
\label{fig1}
\end{figure}

On the other hand, the free energy for free pairs of HQVs 
with their distance $R ( < 2a) $ is obtained as follows.
\begin{eqnarray}
F_{fH}& =&\frac{\pi}{2} \chi_N C^2 \left\{ K
{\rm ln} \left( \frac{a+\sqrt{a^2-R^2/4}}{2 \xi} \right)
+ {\rm ln}(R/\xi) \right. \nonumber\\
& & \left. -\frac{R}{4a} \arcsin \left( \frac{R}{\sqrt{a^2+R^2/4}}
\right)  \right\}
\label{eq_f_fh}
\end{eqnarray}
To get the above free energy, we used the following forms for $\Phi$
and ${\hat d}$.
\begin{equation}
\Phi =\frac{1}{2}
\left[ \tan^{-1} \left( \frac{x+R/2}{y} \right)
+ \tan^{-1} \left( \frac{x-R/2}{y} \right) \right]
\end{equation}
and
\begin{equation}
{\hat d} = {\hat x} \cos{\alpha} + {\hat y} \sin{\alpha}
\end{equation}
with 
\begin{equation}
 \alpha =\frac{1}{2}
\left[ \tan^{-1} \left( \frac{x+R/2}{y} \right)
- \tan^{-1} \left( \frac{x-R/2}{y} \right) \right]
\end{equation}

As in the case of the singular vortex, this solution has ${\hat d}_z =0$.
If we neglect the last term in the Eq. (9), $F_{fH}$ can be 
easily optimized for $R/(2 a) = \sqrt{2 K +1}/(K+1)$ and
\begin{eqnarray}
F_{fH} &\approx& \frac{\pi}{2} \chi_N C^2 \left\{ K {\rm ln}\left(
\frac{a}{2\xi} \frac{1+2K}{1+K} \right)+{\rm ln} \left(\frac{2a}{\xi}
\frac{\sqrt{1+2K}}{2 (1+K)} \right) \right. \nonumber\\
& & \left. -\frac{ 2 \sqrt{1+2K}}{1+K}
\arcsin\left( \frac{\sqrt{1+2K}}{\sqrt{2+4 K+ K^2} } \right) \right\}.
\label{ffh}
\end{eqnarray}
Now we compare the free energies of singular vortex, Eq (\ref{eq_f_singular})
 and 
pairs of free HQV, Eq. (\ref{ffh}), and find that the free HQVs are stable than
the singular vortices if $a/\xi$ satisfies the following equation.
\begin{equation}
\frac{a}{\xi} \ge \frac{1+2K}{2(1+K)} \left( \frac{(1+ 2K)^{3/2}}{(1+K)^2}
\right)^{\frac{1}{(1-K)}}
\end{equation}
In the above equation, the last term in Eq. (\ref{ffh}) is neglected for the 
analytic solution.\cite{footnote}  However, we find that this approximation does not
make a qualitative difference for 
the phase diagram as a function of $(a/\xi)$ and  $K$ shown in Fig 1.
The free HQV is stable than the singular
vortex for all temperatures
( $ 1 < K < 3.295$), since we are interested in small
rotations of $\Omega << \Omega_u$, where  $\Omega_u
\sim 5 \times 10^6$ rad/s which corresponds to $a \sim \xi \sim 0.01 \mu$m.
Hereafter, we will limit ourselves to HQVs and compare
the free energies between free HQVs and bound HQVs.

\section{Free energy of the bound HQVs pairs}
If we allow ${\hat d}$-vector to bend out of the $x-y$ plane
in the vicinity of ${\hat d}$-digyrations, the system will
lower the energy. So it is quite possible that two HQVs
are bound when they are close enough. 
Let us consider the following ${\hat d}$ configurations.
\begin{equation}
{\hat d} = -\cos{\alpha} ({\hat x} \cos{\beta} + {\hat y} \sin{\beta}) + \sin{\alpha} {\hat z}
\end{equation}
where $\beta$ is the cyclic variable, and an example of ${\hat d}$ configurations
is shown in Fig. 2 for $\beta = 0$.

\begin{figure}[t]
\includegraphics[width=2.7in,clip]{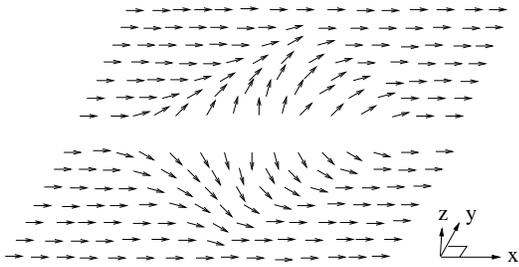}
\caption{A configuration of ${\hat d}$-vector for $\beta=0$ }
\label{fig2}
\end{figure}

Using the free energy formula for the bound HQVs, Eq. (8) in [\onlinecite{kee00}],
we find
\begin{eqnarray}
&F_{bH} & = \frac{\pi}{2} \chi_N C^2 \left\{ K {\rm ln}\left(
\frac{a+\sqrt{a^2-R^2/4}}{2 \xi} \right) + {\rm ln}\left(
\frac{R}{\xi} \right) \right. \nonumber\\
&  & \left. \hskip -1cm -\frac{R}{4a} \arcsin \left(
\frac{R}{\sqrt{a^2+R^2/4}} \right) +
\left( \frac{R}{2\xi_H} \right)^2 {\rm ln}\left(\frac{4 \xi_H}{R}
\right) \right\}
\label{eq_f_bh}
\end{eqnarray}
The free energy can be optimized for $R/a$ for given $K$ and $a/\xi_H$ assuming
that $a/\xi_H >> 1$.
We found that $R/a =0.3$ for $a/\xi_H=10$ for all $K$ values,
and $R/a= 0.15$ for $a/\xi_H = 20$ for all $K$ values. 
For $a/\xi_H =30$, $R/a = 0.1$ for all $K$ values. 
The ratio, $R/\xi_H  (= R/a \times a/\xi_H) \sim 3.0$ 
is independent of $K$ and $a/\xi_H$.
It is remarkable that $R/\xi_H$ is independent of temperatures.

Using the optimal value of $R/\xi_H$, the free energy of the bound HQV can be written as
\begin{eqnarray}
F_{bH} &=&\frac{\pi}{2} \chi_N C^2 \left\{ 
K {\rm ln} \left(\frac{a}{\xi}\right) +
{\rm ln} \left( 3 \frac{\xi_H}{\xi} \right)  \right.
\nonumber\\
& & \left. -\frac{3}{4} \frac{\xi_H}{a} \arcsin\left( \frac{3\xi_H}{a \sqrt{1+
\frac{9\xi_H^2}{4 a^2}}} \right) +\frac{9}{4} {\rm ln}\left(\frac{4}{3}
\right) \right\}.
\label{fbh}
\end{eqnarray}
We compare the free energies of bound HQV, Eq. (~\ref{fbh})
 and free HQV, Eq.  (~\ref{ffh}) neglecting the correction terms of
${\hat d}$ textures (the terms with $\arcsin(..)$).
We find that the bound HQVs is stable over the free HQV, if $a/\xi_H$
 satisfies the following equation.
\begin{equation}
\frac{a}{\xi_H} \ge \frac{8}{3} (2)^K (1+K)^{1+K}(1+2 K)^{-(K+\frac{1}{2})}
\end{equation}
The phase diagram for free HQV and the bound HQV phases as a function of
$a/\xi_H$ and $K$ is shown in Fig. 3.
We see that the bound HQV is more stable than the free HQV when
$a \ge 10 \xi_H$ for all temperatures. 
In other words, the bound pair HQVs are always more stable than free HQVs
for relatively small rotation, $\Omega < \Omega_c$,
where $\Omega_c \sim 2$ rad/s,  assuming
that $\xi_H \sim$ 5 $\mu$m.
\begin{figure}[t]
\includegraphics[width=2.7in,clip]{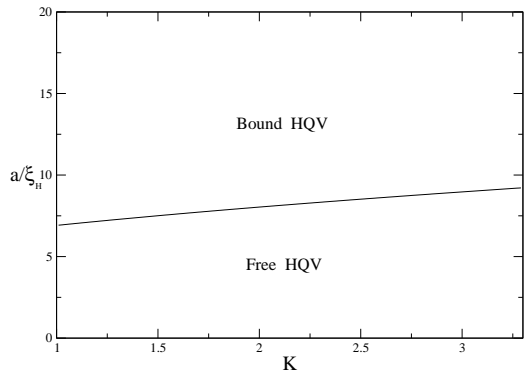}
\caption{Phase diagram for bound HQV phase, and
free HQV phase as functions of $K$ and $a/\xi_H$. Note that the free
HQV is only stable for a relatively small $a/\xi_H$ for all $K$}
\label{fig3}
\end{figure}
From these results, we analyze the $\Omega$ dependence of the satellite
intensity ($I_g$) as follows.

In the small rotation speed, the bound pairs are most stable.
So in the equilibrium condition, the satellite intensity increases
linearly with $\Omega$. When $\Omega$ reaches $\Omega_c$,
the free HQV pairs become more stable. Many bound pairs are converted
into free pairs. This looks a case of phase separation.
In order to clarify this situation, NMR experiment with the field
gradient in the parallel plates is highly desirable.
It is also interesting to note that the $\Omega$ dependence of the satellite
intensity $I_g$ looks surprisingly similar to $-4 \pi M$ versus $B$ 
in typical type II superconductors as first formulated by Abrikosov.
\cite{abrikosov}

\section{Spin wave spectrum}
As we mentioned above, both the singular vortices and free HQVs do not produce
any characteristic NMR signals, since their presence does not disturb
the configuration of ${\hat l} \perp {\hat d}$.
On the other hand, the bound pairs of HQVs are clearly visible by NMR as
discussed in [\onlinecite{kee00,hu87}].
We adopt the formulation for the NMR satellite frequencies
and intensities studied in [\onlinecite{hu87}].
However, in the present case there will be no potential energy for the longitudinal mode,
since the rotation of ${\hat d}$  around the z-axis  does not change the dipole energy.  
In other words, the bound pairs of HQVs as well as singular vortices and free HQVs are 
invisible by  longitudinal NMR. All of them give the no longitudinal resonance. 
In contrast, the transverse NMR is of great interest.  

The transverse resonance frequency, $\lambda_g$  is
determined as follows.
Comparing to [\onlinecite{hu87}], we consider the bound HQVs
when ${\vec H} \parallel {\hat z}$ and $\xi_H < \xi_D$. 
Let us consider a small oscillation of ${\hat d}$ as follows.
\begin{eqnarray}
{\hat d} +\delta {\hat d} &=& \cos{(\alpha+g)} \left[ \cos(\beta+f) {\hat x} +\sin{(\beta+f)} {\hat y} \right]\nonumber\\
& &   + \sin{(\alpha+g)} {\hat z}.
\end{eqnarray}
The change of free energy due to the small oscillation up to the quadratic order
in $f$ and $g$ is found as follows.
\begin{eqnarray}
\delta F = \sum_{i,j} |\nabla_i \delta {\hat d}_j|^2 + \xi_H^{-2} |\delta d_z|^2 
&=& \cos^2{(\alpha)}|{\vec \nabla} f|^2 +|{\vec \nabla} g|^2\nonumber\\
&& \hskip -1.3cm + \;\; \xi_H^{-2} (1-2\sin^2{(\alpha)}) g^2.
\end{eqnarray}
Therefore, the satellite frequencies of $\lambda_f$ and $\lambda_g$ are obtained
by the following eigenvalue equations.
\begin{eqnarray}
-\xi_H^2 \nabla (\cos^2{(\alpha)} \nabla f)  &=& \lambda_f \cos^2{(\alpha)} f \nonumber\\
-\xi_H^2 \nabla^2 g - (1-2 \sin^2{(\alpha)}) g &=& \lambda_g g,
\end{eqnarray}
As we mentioned above, the longitudinal mode has a trivial solution of $f= constant$
with $\lambda_f=0$. Therefore, there will be no longitudinal resonance.
As to the transverse mode, the potential $V_g(x,y) = -1+2 \sin^2{(\alpha)}$ is shown in Fig. 4,
and $\sin^2{(\alpha)}$ is given by 
\begin{equation}
\sin^2{(\alpha)} = \frac{\sin^2{v}}{\sinh^2{u} +\sin^2{v}}
\end{equation}
using the change of parameters as
\begin{equation}
x=\frac{R}{2} \cosh{u} \cos{v}, \;\;\;\;\; 
y=\frac{R}{2} \sinh{u} \sin{v},  
\end{equation}
%
%
\begin{figure}[t]
\includegraphics[width=3.5in,clip]{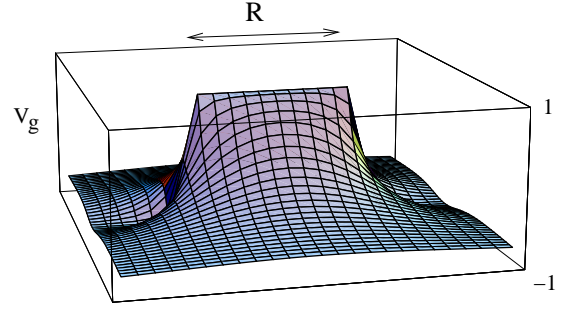}
\vskip -0.83cm
\caption{Potential for the transverse mode is shown in $x-y$ plane. It is
repulsive between two HQVs, where $R$ is
the optimal separation between the bound HQVs. }
\label{fig4}
\end{figure}
Note that the potential energy in the present case is completely different 
from the one considered in [\onlinecite{hu87}].
This implies (a) first  of all spin waves are extended,  and (b) the related intensities are 
 $ \sim \pi a^2 n_v$.  
At the critical rotation of $\Omega=\Omega_c$, $n_v = (\pi a^2)^{-1}$.

These eigenequations can be solved using a variational function.
Note that there is a trivial solution of $g_0 = {\rm sech}{u}$ with $\lambda_g =-1$,
which does not give a distint satellite signal. So we look for
a solution which is orthogonal to $g_0$ and have a node in the spin wave function.
We find the following function of $g$ with a vartional parameter $\mu$
satisfies the above criteria. 
\begin{equation}
g=({\rm sech}{u})^{\nu} - p \times {\rm tanh}{u} \times ({\rm sech}{u})^{\mu},
\end{equation}
where $\nu=1.13$ and $p=(\mu^2-1) (B[\frac{1}{2},\frac{\nu-1}{2}]-\frac{1}{2} 
B[\frac{1}{2},\frac{\nu+1}{2}])/(\mu+3)$, and $B$ is the Euler beta function. 
With the above variational functions, we find
the optimal value of $\lambda_g$ which is obtained by
minimizing $\lambda_g$ with respect to $\mu$. 
$\lambda_g$ vs. $\xi_H/R$ is shown in Fig. 5. 
Note that $R/\xi_H \sim 3$ is independent of $a/\xi_H$ and $K$, as long as the system is in
the regime where the bound HQVs are stable.
The value of $\lambda_g \sim -0.05$ for $R/\xi_H =3$ is close to the satellite
frequency, $\lambda_g = -0.06 \sim -0.09$, observed in [\onlinecite{yamashita}].
On the other hand, we find that the bound state does not exist, when $\xi_H/R > 0.35$. 

\begin{figure}[t]
\includegraphics[width=4.5in,clip]{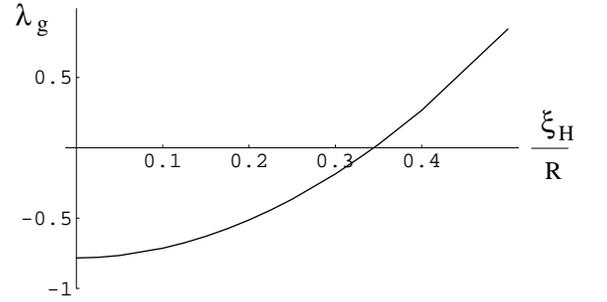}
\vskip -3.8cm
\caption{The satellite frequency of the transverse mode
as a function of $\xi_H/R$ is shown for the bound HQVs. }
\label{fig5}
\end{figure}


The intensity of the transvermode is obtained as
\begin{equation}
I_g = \frac{n_v}{2} \frac{|\int_0^{arccosh(2a/R)} e^{i\alpha} g d u|^2}{ \int_0^{arccosh(2a/R)} |g|^2 d u}.
\end{equation}
For $R/\xi_H=3$, $I_g$ depends on $a/\xi_H$.
We find that  $I_g = 23.4 (R/2)^2 n_v$
and $I_g = 43.2 (R/2)^2 n_v$ for $a/\xi_H=7$, and 
$a/\xi_H = 10$, respectively.
Since $n_v = (\pi a^2)^{-1}$ at $\Omega = \Omega_c$, $I_g$ at the critical rotation is given by
\begin{eqnarray}
I_g & = & \frac{43.2}{\pi} \left( \frac{3}{20} \right)^2 = 0.31 
\;\;\; {\rm for}  \;\;\; a/\xi_H=10,
\nonumber\\
& =& \frac{23.4}{\pi} \left( \frac{3}{14} \right)^2 =0.34
\;\;\; {\rm for}  \;\;\;  a/\xi_H=7.
\end{eqnarray}
The ratio between the satellite intensity and total intensity, $\delta I/I$ in 
the NMR spectra was observed as $\delta I /I =0.2$ at $T =0.8 T_c$,
and $0.1$ at $T=0.93 T_c$.\cite{yamashita}
Therefore,  the observed signal is about 60 \%  and 30 \% of the theoretical value at $T=0.8 T_c$
and $T=0.93 T_c$, respectively.
In our analysis, $I_g$ does not explictily depend on temperature, while it is possible that
the vortex pinning changes the satellite intensity. 
Therefore, the temperature dependence in the experiment suggests that the vortex pinning becomes weaker as temperature decreses.

\section{Discussion and Summary}
Since Salomaa-Volovik's proposal for the HQVs in 1985, HQVs have been searched
in the rotating superfluid $^3$He-A in the confined geometry.
However, all of earlier experiments have been done in the wider gap $D$ ($>25 \mu m$). 
Then the most
common texture is the vortex sheet associated with the composite soliton.\cite{heinila95}
In the perspective, the vortices seen in Walmsley et al\cite{walmsley04} appears
to be the vortex sheet as well.
The experimental data in [\onlinecite{yamashita}] is quite different from the earlier ones
in the following ways.
(a) A new transverse satellite with $\lambda_g = -0.06 \sim -0.09$ is observed.
(b) Both the satellite frequency and the critical value of $\Omega_c$ 
where the maximum value of $\delta I/I$ is observed
are almost independent of temperature.
(c) The critical value of $\Omega_1$ where the first appearance of the satellite singal
occurs and the maximal value of $\delta I/I$ appear to weakly 
depend on temperature.

The above observations are in fact consistent with our HQV scenario. 
We have shown that $R/\xi_H \sim 3.0$ is independent of $H$ and $T$.
Also from the universality of this ratio, we predict that the satellite frequency
and the intensity are almost independent of $H$ and $T$, 
when we ignore the vortex pinning effect.
On the other hand, any texture involving ${\hat l}$-texture should depend strongly
on temperature and gap size $D$. 
Furthermore, another satellite frequencies associated with ${\hat l}$-bend textures
should appear near $\lambda_g \sim -1$.\cite{bruinsma79}
Absence of these satellites clearly indicate that the satellites seen in Yamashita et al
\cite{yamashita} should be due to the bound pairs of HQVs.
The parallel experiments with different size of $D$
are highly desirable to distinguish the ${\hat d}$-texture vs. ${\hat l}$-textures
as well,
as our study indicates that the satellite frequency is independent of $D$.

\vskip 1.5cm

\begin{acknowledgments}
We would like to thank Takao Mizusaki and Minoru Yamashita for providing us their data
on the ongoing superfluid $^3$He experiments at ISSP, University of Tokyo at Kashiwanoha and
their useful correspondences, which motivated the present study.
This work was supported by NSERC of Canada, Canada Research Chair,
Canadian Institute for Advancded Research, and Alfred P. Sloan
Research Fellowship(HYK).
One of us (KM) thanks Hyekyung Won at Hallym University for hospitality, where a part of 
the work was carried out. 
\end{acknowledgments}


\begin{thebibliography}{100}
\bibitem{footnote} Note when $K$ approaches to $1$, $a/\xi$ approaches to $0$, since $1/(1-K)$ becomes $-\infty$
\end{thebibliography}
\end{document}